\documentclass{iopart}

\usepackage{graphicx}
\usepackage{amssymb}

\newcommand{\beq}{\begin{equation}}
\newcommand{\eeq}{\end{equation}}

\newcommand{\pd}{\partial}
\newcommand{\F}{{\cal F}}
\newcommand{\C}{{\cal C}}

\newcommand{\MeV}{\,\mathrm{MeV}}

\newcommand{\Hz}{\,\mathrm{Hz}}
\newcommand{\nHz}{\,\mathrm{nHz}}

\newcommand{\Tdec}{T_\mathrm{dec}}

\begin{document}

\title[Signature of neutrino decoupling in the nHz region of the GW spectrum]{A possible signature of cosmic neutrino decoupling in the nHz region of the spectrum of primordial gravitational waves}

\author{Massimiliano Lattanzi$^{1}$, Riccardo Benini$^1$ and Giovanni Montani$^{1,2}$ 
}
\address{$^1$ ICRA and Physics Department, University of Rome, ``La Sapienza'', P.le Aldo Moro 2, 00185 Rome, Italy}
\address{$^2$ ENEA -- C.R. Frascati (Department F.P.N.), Via Enrico Fermi, 45 (00044), Frascati (Rome), Italy}
\ead{lattanzi@icra.it}

\begin{abstract}
In this paper we study the effect of cosmic neutrino decoupling on the spectrum of cosmological gravitational waves (GWs). At temperatures $T\gg 1\MeV$,
neutrinos constitute a perfect fluid and do not hinder GW propagation, while for $T\ll1\MeV$ they free-stream and have an effective viscosity
that damps cosmological GWs by a constant amount. In the intermediate regime, corresponding to neutrino decoupling, the damping is frequency-dependent.
GWs entering the horizon during neutrino decoupling have a frequency $f\sim 1\nHz$, corresponding to a frequency region that will be probed by Pulsar Timing Arrays (PTAs). In particular, we show how neutrino decoupling induces a spectral feature
in the spectrum of cosmological GWs just below 1 nHz. We briefly discuss the conditions for a detection of this feature and conclude that it is unlikely to be
observed by PTAs.
\end{abstract}

\pacs{04.30.Nk, 95.85.Sz, 14.60.Lm}

\section{Introduction}
\label{intro}

The presence in the Universe today of a stochastic background of gravitational waves (GWs) is a quite general prediction of several early cosmology scenarios. In fact, the production of gravitational waves is the outcome of many processes that could have occurred in the early phases of the cosmological evolution, like the amplification of vacuum fluctuations in inflationary \cite{Gr75} and pre-big-bang cosmology scenarios \cite{Ga93}, phase transitions \cite{Ho86}, and finally the oscillation of cosmic strings loops \cite{Vi81}. In most of these cases, the predicted spectrum of gravitational waves extends over a very large range of frequencies.

The detection of such primordial gravitational waves, produced in the early Universe, would be a major breakthrough in cosmology and high energy physics. This is because gravitational waves decouple from the cosmological plasma at very early times, when the temperature of the Universe is of the order of the Planck energy. In this way, relic gravitational waves provide us a ``snapshot'' of the Universe near the Planck time, in a similar way as the cosmic microwave background  (CMB) radiation images the Universe at the time of recombination.

The extremely low frequency region ($f\lesssim10^{-15}\Hz$) in the spectrum of primordial gravitational waves can be probed through the anisotropies of the CMB. In particular, gravitational waves leave a distinct imprint in the so-called magnetic or B-modes of its polarization field \cite{Hu97,Pr05}. 
The frequency region between $\sim 10$ Hz and few kHz is probed by operating large scale interferometric GW detectors like LIGO \cite{LIGO} and VIRGO \cite{VIRGO}, that, although designed with the aim to detect astrophysical signals, can possibly also detect signals of cosmological origin \cite{Ma00}. 
The LISA space interferometer \cite{LISA}, that will hopefully operate in the 2020s, will probe the frequency region between $10^{-4}$ and $1\,\Hz$.
Finally, pulsar observations can be used to obtain information on the stochastic GW background, through the technique known as pulsar timing. The so-called ``Pulsar Timing Arrays'' will probe the region from $\sim$1 to $\sim$ 100 nHz, with a maximum sensitivity between $\sim$3 and $\sim$10 nHz \cite{PTA}.

In order to compare the theoretical predictions with the expected instrument sensitivities, one needs to evolve the GWs from the time of their production to the present. 
This is done by assuming that gravitational waves propagate freely across the Universe. In this case, the only effect is that the amplitude of the wave decreases due to the expansion of the Universe. However, GWs are sourced by the anisotropic stress part of the energy-momentum tensor of matter, so that the above assumption is well-motivated only when anisotropic stress can be neglected. It is already known that the anisotropic stress of free streaming neutrinos acts as an effective viscosity, absorbing gravitational waves in the low frequency region, thus resulting in a damping of the B-modes of CMB \cite{Bo96,Du98,We04,Xi08,Zh09}.\footnote{ In Ref. \cite{Mangilli:2008bw}, the evolution of cosmological GWs in the presence of neutrino free-streaming was studied up to second order in perturbation theory. Moreover, GW propagation is also affected by entropy-generating events (like the $e^+e^-$ annihilation) occurring during the cosmological evolution; the effect on the GW spectrum has been studied in Ref.  \cite{Watanabe:2006qe}. The authors of Ref. \cite{Watanabe:2006qe} also consider the effect of neutrino free streaming but assume an instantaneous neutrino decoupling. 
A similar analysis, including an accurate computation of the primordial spectrum as generated during inflation, and extended to higher frequencies, can be found in Ref. \cite{Kuroyanagi:2008ye}.
Finally, in Ref. \cite{Ichiki:2006rn}, the role of a non-vanishing lepton asymmetry in determining the amount of absorption has been considered. (\emph{Note added in the arXiv version.})}

In the present work we aim to study a possible signature of neutrino decoupling on the spectrum of cosmological GWs. During decoupling, the effective viscosity
of neutrinos is increasing from zero (its value at high temperatures, when the neutrinos are tightly coupled to the cosmological plasma) to a finite value.
Neutrino decoupling happens when the temperature of the Universe is $T\simeq$ 1 MeV, or redshift $z\simeq 10^{10}$; GWs entering the horizon at that time
have a frequency $f\simeq 1\nHz$. Since the viscosity results in a damping of the wave, we expect to have a frequency-dependent absorption of GWs in this
frequency range. A spectral feature in the nHz range would be potentially interesting for PTAs. PTAs combine the fact that pulsars are very stable clocks with the fact that the time of arrival of the pulse also depends on the GW background between the pulsar and the Earth \cite{PTA}. The use of a pulsar array allows to correlate the electromagnetic signals from different pulsars, thus eliminating variations in the arrival times that do not depend on the GW background. The dominant contribution to the GW background in the PTAs sensitivity range is that of supermassive black hole binaries following galaxy mergers \cite{Ja03,Jenet:2005pv,Sesana:2008mz}. This has an astrophyisical origin and would obviously not be affected by the decoupling of cosmological neutrinos. The dominant \emph{cosmological} constribution to the GW signal in the nHz range is expected to be
that of cosmic strings, that are topological defects left after symmetry-breaking phase transitions occurring in the early Universe. Other, possibly important, contributions in this frequency range are those of cosmic superstrings (i.e., topological defects arising in string-theory inspired inflationary models) and of
inflation-generated GWs (through the usual amplification of quantum zero-point fluctuations). For our purposes, one important difference between the cosmic (super)strings-generated GWs on one side, and the inflationary GWs on the other, is that the former are created by causal processes at horizon and subhorizon scales, while the latter are generated by the stretching of quantum fluctuations to super-horizon scales. In the following we will concentrate on GWs that were outside the horizon at some time before neutrino decoupling, so that our results rigorously applies to inflationary GWs but not necessarily to string-generated GWs. It should be noted that the inflationary signal is, in the nHz range, the weakest of the cosmological signals mentioned above, and is probably beyond the reach of the ongoing PTA projects. It could be however within the reach of future instruments  \cite{PTA,Jenet:2009hk} like the Square Kilometer Array (SKA)   \cite{Kramer:2004hd}.

The paper is organized as follows. In Sec. 2, we recall the basic equations describing the evolution of GWs in the presence of an imperfect fluid. 
In Sec. 3, we discuss the interaction between GWs and the neutrino fluid in the three regimes $T\gg 1\MeV$, $T\simeq 1\MeV$ and $T\ll 1\MeV$, corresponding to
$f\gg 1\nHz$, $f\simeq 1\nHz$ and $f\ll 1\nHz$ respectively. In Sec. 4, we present the results of the numerical integration of the Einstein-Boltzmann system, 
show the modification in spectrum of cosmological GWs and briefly discuss the conditions that could lead to its detection.

\section{Basic equations}
\label{sec:prop}

We shall use, all throughout the paper, natural units in which $c=\hbar=k_B=1$.

Let us consider a gravitational wave, propagating on the background of a flat (k=0) Friedmann Universe. In synchronous gauge, the spatial components of the perturbed metric are written as
[we use the $(-+++)$ signature for the metric]
\begin{equation}
g_{ij}=a^2(t)\left[\delta_{ij}+h_{ij}\right]
\end{equation}
while the other components are left unperturbed: $g_{00}=-1$ and $g_{0i}=0$. We will consider only the transverse traceless part of $h_{ij}$. With this restriction,
$h_{ij}$ only has two degrees of freedom, corresponding to the two polarizations of a GW. 
Here $a(t)$ is the cosmological scale factor, that evolves according to the background Friedmann equation:
\begin{equation}
\left(\frac{da}{dt}\right)^2=\frac{8\pi G}{3} a^2 \bar\rho
\label{eq:fried}
\end{equation}
where $\bar \rho$ is the background density of the cosmological fluid (in general, we will use overbars to denote background quantities).

The Einstein equation for the evolution of a transverse traceless metric perturbation $h_{ij}$ takes the form of a wave equation \cite{We04,We72,We08}:
\begin{equation}
\partial^2_t{h}_{ij}+\left(\frac{3}{a}\frac{da}{dt}\right)\partial_t{h}_{ij}-\left
(\frac{\nabla^2}{a^2}\right)h_{ij}=16\pi G\frac{\Pi_{ij}}{a^2}\;,
\label{eq:einst}
\end{equation}
where $\Pi_{ij}$ is the anisotropic stress, i.e. the traceless part of the three dimensional energy-momentum tensor $T^{i}_{j}$ of the cosmological fluid, representing
dissipative effects that are not present in a perfect fluid. It is then defined through the relation $\Pi^i_j=T^i_{ j}-\delta^i_j T_k^k/3=T^i_j-{\cal P}\delta^i_{j}$, where $\cal{P}$ is the total pressure of the fluid (including possibly a small perturbation with respect to the background). We assume that the anisotropic stress is relevant only for one component of
the cosmological fluid (the neutrinos), providing a density $\bar\rho_\nu =f_\nu \bar\rho$, while for the other components (e.g., the photons) it can be safely neglected.

In the following, we will find more convenient to use the conformal time $\eta$, related to the synchronous time $t$ by $dt=a(\eta)d\eta$ as our time variable. 
Moreover, since the perturbation equations are linear, it is very convenient to Fourier transform the spatial dependence of all the relevant quantities. Then the 
wave Eq. (\ref{eq:einst}) rewrites as:
\begin{equation}
\ddot{h}_{ij}+2\mathcal{H}\dot{h}_{ij} + k^2 h_{ij}=16\pi G\Pi_{ij}\;,
\label{eq:einstK}
\end{equation}
where dots denote derivatives with respect to $\eta$, and $\mathcal{H}$ is the conformal Hubble constant $\mathcal{H} = \dot a/a$.
For the sake of simplicity, we use the same symbol for a given quantity and for its Fourier transform.

In a perfect fluid, anisotropic stress is absent by definition (in fact, for a perfect fluid $T^i_j = \mathcal{P}\delta^i_j$), so that Eq. (\ref{eq:einst})
takes the homogeneous form:
\begin{equation}
\ddot{h}_{ij}+2\mathcal{H}\dot{h}_{ij} + k^2 h_{ij}=0.
\label{eq:einstK0}
\end{equation}
In the radiation-dominated era, $a\propto \eta$ and $\mathcal{H}=1/\eta$. Then, considering the quantity $\chi_{ij}\equiv k\eta  h_{ij}$ we find that 
\begin{equation}
\ddot \chi_{ij} + k^2 \chi_{ij} =0.
\end{equation}
This is a simple, undamped oscillator equation, that admits the solutions $\chi_{ij}^{(1)} = \sin k \eta$ and $\chi_{ij}^{(2)} = \cos k\eta$, i.e.:
\begin{equation}
h_{ij} = a_{ij}\frac{\sin (k\eta)}{k\eta} + b_{ij} \frac{\cos (k\eta)}{k\eta}.
\end{equation}
In the limit of $k\eta \to 0$, i.e., for a wave that is far outside the horizon, we have that
\begin{equation}
h_{ij} = a_{ij} + \frac{b_{ij}}{k\eta}.
\end{equation}
Then, the most general solution outside the horizon (during the radiation dominated era) is the superposition of a constant mode and of a mode that decays like $1/\eta$.
This is actually true also in the presence of anisotropic stress, because dissipative effects cannot operate on scales larger than the horizon, so that
$\Pi^i_j$ has to vanish in the limit $k\eta \to 0$.

The macroscopic properties of the cosmological fluid, including the anisotropic stress, can be derived by the phase space distribution of its particles. The phase space is described by three positions $x^i$ and by their three conjugate momenta $P_i\equiv mdx_i/ds$, 
and evolves according to the Boltzmann equation:
\begin{equation}
\hat L[f]\equiv\frac{df}{d\eta}=\frac{d x^\mu}{d \eta}\frac{\pd f}{\pd x^\mu}+\frac{dP^\mu}{d\eta}\frac{\pd f}{\pd P^\mu}=\hat C[f]
\label{eq:boltzeq}
\end{equation}
where $\hat L\equiv d/d\eta$ is called Liouville operator, and $\hat C$ is the collision operator accounting for changes in the distribution function due to collisions between particles. Using the geodesic equation, the total derivative of $f$ can be rewritten as:
\begin{equation}
\frac{df}{d\eta}=\frac{d x^\mu}{d \eta}\frac{\pd f}{\pd x^\mu}-\Gamma^\mu_{\alpha\beta}P^\alpha P^\beta\frac{\pd f}{\pd P^\mu}.
\label{eq:liouv}
\end{equation} 
The link between the Einstein Eq. (\ref{eq:einst}) describing the evolution of the metric perturbation $h_{ij}$ and the Boltzmann Eq. (\ref{eq:boltzeq}) is given 
by the following expression for the energy-momentum tensor in terms of the distribution function:
\begin{equation}
T^\mu_\nu=\frac{1}{\sqrt{-g}}\int f(x^i, P_j,t) \frac{ P^\mu P_\nu}{P^0}\,dP_1\,dP_2\,dP_3,
\label{eq:Tmunu}
\end{equation}
where $g$ is the determinant of the metric. Then, once the collision term is also specified, Eqs. (\ref{eq:fried}), (\ref{eq:einst}),  (\ref{eq:boltzeq})  and (\ref{eq:Tmunu}) are all that is needed, at least in principle, to follow the propagation of a GW.

In general, the Einstein-Boltzmann system for the coupled evolution of $h_{ij}$ and $f$ is an integro-differential system, because the source term in Eq. (\ref{eq:einst}) is given by an integral
over $f$. The system can be however reduced to an ordinary differential system in the case of massless particles. This is made by integrating out the dependence
of $f$ on energy and expanding its dependence on the direction of the particle momentum in Legendre polynomials. This method is widely applied
in the numerical treatment of scalar perturbations \cite{Ma:1995ey}. The analogous equations for the tensor case have been derived in Ref. \cite{BLM10}, 
to which we refer the reader for further details; here we just quote the relevant equations. The coupled Einstein-Boltzmann system is equivalent
to the following system of infinite differential equations in $k$-space:

\begin{eqnarray}
\fl \ddot h_{ij} + 2{\cal H} \dot h_{ij} + k^2 h_{ij} =4 G a^2 \bar \rho_\nu \mathcal{F}_{ij}^{(0)} \label{eq:hij},\\[0.3cm]
\fl \dot \F_{ij}^{(0)} = - k\,\F_{ij}^{(1)} - \frac{8\pi }{15} \dot h_{ij } +\C_{ij}^{(0)}, \label{eq:bolt_mpol1} \\[0.3cm]
\fl \dot \F_{ij}^{(2)} = -\frac{k}{5} \left[   3 \F_{ij}^{(3)} -  2 \F_{ij}^{(1)} \right] - \frac{16\pi}{105} \dot h_{ij }+\C_{ij}^{(2)},\\[0.3cm]
\fl \dot \F_{ij}^{(4)} = - \frac{k}{9} \left[   5 \F_{ij}^{(5)} -  4 \F_{ij}^{(3)} \right] - \frac{8\pi}{315} \dot h_{ij } +\C_{ij}^{(4)},\\[0.3cm]
\fl \dot \F_{ij}^{(\ell)} = -\frac{k}{2\ell+1} \left[   (\ell+1) \F_{ij}^{(\ell+1)} -  \ell\, \F_{ij}^{(\ell-1)} \right] +\C_{ij}^{(\ell)}\qquad (\ell \ne 0,2,4). \label{eq:bolt_mpol4}
\end{eqnarray} 

In the above equations, the $\mathcal F^{(\ell)}$'s and the $\mathcal C^{(\ell)}$'s are related to the coefficients of the expansion in Legendre
polynomials of $f$ and $\hat C[f]$, respectively (see Ref. \cite{BLM10} for details). The anisotropic stress is given by:
$\Pi_{ij}=a^2\bar\rho_\nu\F_{ij}^{(0)}/4\pi$.

\section{Interaction of gravitational waves with neutrinos}

In this section we discuss how cosmological neutrinos affect the propagation of cosmological GWs. 
We restrict our attention to waves entering the horizon well
before the time of matter-radiation equality, corresponding to a redshift $z\simeq 10^4$. This corresponds
to waves with a present frequency $\nu \gg 10^{-16}$ Hz. We recall that in the radiation-dominated era, $a\propto \eta$ and ${\cal H}=1/\eta$.

The thermal evolution of neutrinos can be divided into two distinct regimes. In the early Universe, when the temperature
was sufficiently high, neutrinos were kept in thermal equilibrium with the cosmological plasma by frequent reactions
like $\nu\bar \nu\leftrightarrow e^+ e^-$, $\nu e\leftrightarrow \nu e$, etc. As the Universe expanded and cooled down.
the interaction rate $\Gamma_\nu$ for these reactions eventually became smaller than the expansion rate $H$ and the neutrinos 
decoupled from the cosmological plasma. From this point on, neutrinos can be considered as freely streaming through the Universe.
The temperature $\Tdec$ that separates the collisional from the non-collisional regime, defined as the temperature
when the interaction rate is equal to the  expansion rate, i.e. $\Gamma_\nu(\Tdec)\simeq H(\Tdec)$,
is found to be $\Tdec\simeq 1 \MeV$, corresponding to $z\simeq 10^{10}$ and $t\sim$ a few seconds.
A wave that enters the horizon at that time has a frequency $f \sim 1 \nHz$. Higher frequency waves enter
the horizon \emph{before} neutrino decoupling, and viceversa.
Then, the following three regimes for the evolution of a GW can be correspondingly identified:
\begin{itemize}
\item When $T\gg 1\MeV$, the neutrinos are tightly coupled to the cosmological plasma, that 
behaves like a single, perfect fluid. The anisotropic stress of neutrinos (as well as of the other components)
is negligible and the GW evolves according to the homogeneous Eq. (\ref{eq:einstK0}). 
This regime is relevant for waves with $f\gg 1\nHz$.
\item When $T\simeq 1\MeV$, the neutrinos are decoupling from the plasma. They cannot be considered
as a perfect fluid because the mean free path of particles is getting large, so that the anisotropic stress cannot be neglected. 
However, the collisions have still to be taken into account. The GW evolution has to be calculated from the full system Eqs. (\ref{eq:hij})-(\ref{eq:bolt_mpol4})
with the $\mathcal{C}^{(\ell)}$'s modeling the collision processes. This regime is relevant for waves with $f\simeq 1\nHz$.
\item When $T\ll 1\MeV$ the neutrinos are collisionless and behave as free particles. The anisotropic
stress cannot be neglected: the GW evolution has to be calculated from the full system Eqs. (\ref{eq:hij})-(\ref{eq:bolt_mpol4})
with $\mathcal{C}^{(\ell)} = 0$ for all $\ell$'s. This regime is relevant for waves with $f\ll 1\nHz$.
\end{itemize}

In the first, high-temperature regime, the metric perturbation simply evolves as $h_{ij}=\bar h_{ij}\equiv\lambda_{ij}\sin(k\eta)/k\eta$ 
(assuming the initial velocity vanishes), where $\lambda_{ij}$ is the initial value of the perturbation. 
In other words, the GW oscillates with a time-dependent amplitude $\propto 1/k\eta$. This can be seen
as the energy loss of the wave due to the expansion of the Universe.

The third, low-temperature regime was first studied in Ref. \cite{We04}, where it was realized that 
the anisotropic stress of neutrinos would lead to a partial absorption of the GW.
In particular, for the standard value of the neutrino density, it was found that the amplitude of the wave when it is
well inside the horizon is reduced to 80\% of its value in the absence of stress. The intensity is reduced to 
$(0.8)^2=64\%$ of its value in the absence of stress. In Ref. \cite{We04}, the Boltzmann-Einstein
system was treated in its integro-differential form. We re-analyzed the problem in Ref. \cite{BLM10} using
the multipole formalism, and allowing for different values of the neutrino density. In general, one has that the
behaviour of the wave well inside the horizon ($k\eta \gg 1$) is of the kind:
\begin{equation}
h_{ij} = D(f_\nu) \lambda_{ij}\frac{\sin(k\eta+\phi)}{k\eta} \qquad(k\eta\gg 1),
\end{equation}
where $D$ is a damping factor, $\phi$ is a phase and $f_\nu\equiv \bar \rho_\nu /\bar \rho$ is the neutrino fraction, i.e.
the ratio of the neutrino density $\bar \rho_\nu$ to the total density $\bar \rho$. This is constant in the radiation-dominated
era since both $\bar \rho_\nu$ and $\bar \rho$ scale like $a^{-4}$, and is equal to $f_\nu =0.4052$ for the standard case of three neutrino families with
a present temperature $T^0_\nu = (4/11)^{1/3} T^0_\gamma=1.9$~K. In the following we will neglect the phase shift $\phi$,
since we are only interested in the behaviour of the wave when averaged over many periods. The damping factor
$D$ only depends on $f_\nu$. The result of Ref. \cite{We04} can thus be written as $D(f_\nu = 0.4052)=0.8$.
In Ref. \cite{BLM10} we have found the slightly larger value $D(f_\nu = 0.4052)=0.88$. We have also calculated
the maximum amount of damping, corresponding to $D(f_\nu=1)=0.75$. It is worth stressing that the cosmological
neutrino background has not been directly observed yet. Although strong deviations from the standard scenario 
with $f_\nu=0.4052$ are unlikely, the possibility that $f_\nu$ has a different value
(due for example to the presence of additional free-streaming particles in the early Universe) should be taken into account.

It is worth stressing, at this point, that the behaviour of the wave in either of the two extreme regimes $f\gg 1 \nHz$ and $f\ll 1 \nHz$
does not depend on the frequency of the wave nor on the values of the cosmological parameters (at least for waves entering the horizon
before matter-radiation equality). This is due to the fact that, when $\mathcal{C}^{(\ell)}=0$ and $a\propto \eta$ 
the explicit $k$ dependence in the Einstein-Bolztmann system can be eliminated by considering the time variable $u=k\eta$. Also,
the convenience of using $u$ is that $u=k\eta \sim 1$ corresponds to the time of horizon crossing. 
Moreover, by writing $h_{ij}=\lambda_{ij}\bar h(\eta)$, it can be seen that $h(\eta)$ satisfies the same differential equations
as $h_{ij}$ [upon a redefinition of the $\mathcal{F}^{(\ell)}$'s and $\mathcal{C}^{(\ell)}$'s], so that we can always take
without loss of generality $\lambda_{ij}=1$. Taking all this into account, we have that the GW evolution for $k\eta\gg 1$, in terms of $\chi_{ij}=k\eta h_{ij} = u h_{ij}$
is:
\begin{equation}
\chi_{ij}(u) = \left\{
\begin{array}{ll}
\sin(u) & (f\gg 1 \nHz)\\[0.2cm]
D(f_\nu)\sin(u) & (f\ll 1 \nHz)
\end{array}
\right.
\end{equation}

The fact that the (rescaled) amplitude of the wave is going from $1$ for $f> 1\nHz$ to $D(f_\nu)<1$ for $f< 1\nHz$ already points
out to the fact that one should expect a frequency dependence in the region $f\simeq 1\nHz$. In order to quantify this,
one should solve the evolution equations for waves entering the horizon in the intermediate regime where neutrinos are neither a perfect 
fluid nor can be considered as freely streaming. The effect of collisions has to be taken into account. 
The exact computation of the collision term $C[f]$ (and consequently of the $\mathcal{C}^{(\ell)}$'s) depends on the details of the interaction. However, a useful although rough approximation consists in writing $\hat C[f] = - \delta f /\tau$, where $\delta f=f-f^{(0)}$ is the deviation of the neutrino distribution
function $f$ from its thermal equilibrium value $f^{(0)}$, 
and $\tau$ is a characteristic time of the interactions that mantain the equilibrium. This form, albeit very simple, captures 
the main features that we expect from the actual collision term. When the characteristic time of the interactions is very large, $\tau\to\infty$ and
$\hat C[f]\to 0$, i.e. the fluid is collisionless. On the contrary, when the characteristic time is very small $\tau\to 0$ and $\hat C[f]$ is very large
unless $\delta f\to 0$ too, i.e. the frequent collisions tend to mantain the equilibrium. This form for $\hat C[f]$ gives \cite{BLM10}:
\begin{equation}
\mathcal{C}_{ij}^{(\ell)}=-\F_{ij}^{(\ell)}/\tau.
\label{eq:coltau}
\end{equation}

In our calculations, we take $\tau$ to be the mean time between collisions, given by $\tau=(n\langle \sigma v\rangle)^{-1}$, where
$n$ is the number density of particles and $\langle\sigma v \rangle$ is the thermally averaged cross section times velocity.
The number density of particles in thermal equilibrium is $n\simeq T^3$, while an adequate value for the cross section
for weak interaction processes is $\sigma\simeq G_F^2T^2$, where $G_F$ is the Fermi constant. Then:
\begin{equation}
\tau =\frac{1}{n\langle\sigma v\rangle}\simeq \frac{1}{G_F^2 T^5}.
\label{eq:tauT}
\end{equation}
\section{Results and discussion}
\begin{center}
\begin{figure}[ht]
\includegraphics[width=0.9\textwidth]{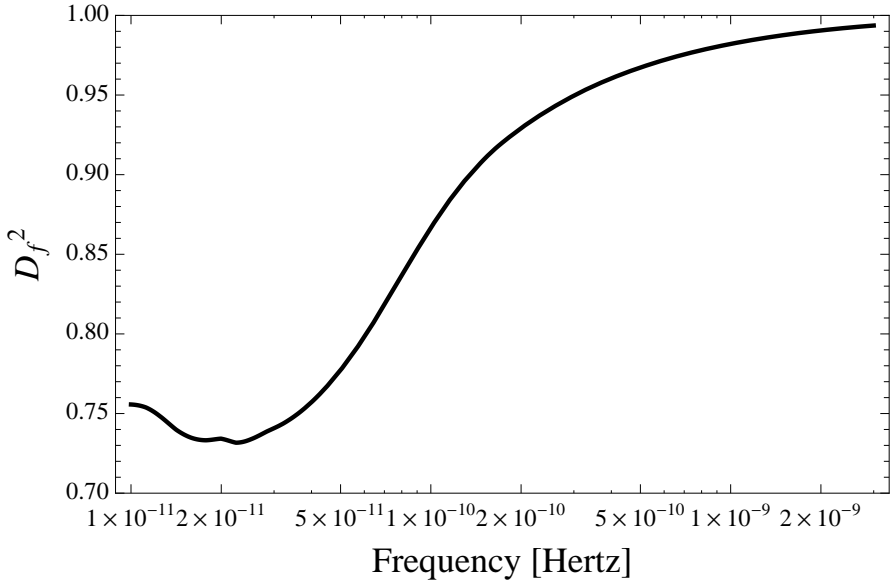}
\caption{The correction factor $D^2_f=\Omega_\mathrm{GW}/\bar \Omega_\mathrm{GW}$. The highest frequencies in the plot
correspond to waves entering the horizon when the neutrinos are tightly coupled to the plasma, so that there is no dissipation.
The lowest frequencies correspond to waves entering the horizon when the neutrinos are freely streaming, so 
that $D^2_f$ is constant and roughly equal to 0.75.}
\label{fig:D2f}
\end{figure}
\end{center}

Equipped with the form (\ref{eq:coltau}) for the collision terms, with $\tau$ given by Eq. (\ref{eq:tauT}), we have integrated the Boltzmann-Einstein system for 71 values
of $f=k/2\pi$ between $10^{-12}\Hz$  and $10^{-8.5}\Hz\simeq 3 \nHz$, with a constant logarithm spacing $\Delta\log f = 0.05$. We 
have assumed the standard value $f_\nu = 0.4052$. For each value of $f$, we have defined a frequency-dependent damping factor $D_f$ as:
\begin{equation}
D_f= \sqrt{2\langle (k\eta h_{ij})^2\rangle}
\end{equation}
where the brackets $\langle\dots\rangle$ denote the average over many periods, taken when the wave is well inside the horizon ($k\eta\gg 1$).
It is clear that when $h_{ij}=\sin(k\eta)/k\eta$ we have $D_f=1$. The damping factor basically quantifies how much the
amplitude of a GW produced in the early Universe would be reduced due to the neutrino anisotropic stress, with respect
to propagation in a perfect fluid. We express our results in terms of the
density parameter of GWs $\Omega_\mathrm{GW}(f)=(d\rho_\mathrm{GW}/d\ln f)/\rho_c$, where $\rho_\mathrm{GW}$ is the energy density of GWs
and $\rho_c$ is the critical density of the Universe.  The quantity $\Omega_\mathrm{GW}$ is used by theorists to quantify the intensity
of the cosmological GW background predicted in a given scenario, and experimental sensitivities are also often quoted in terms of it.
The present intensity is usually computed simply by rescaling the GW amplitude at the source by a factor $1/k\eta$ to take into account
the redshift due to the expansion. This should be corrected to take into account also the anisotropic stress of neutrinos. Since
$\rho_\mathrm{GW} \propto h_{ij}^2$, the correction factor is simply $D_f^2$. Thus, if we call $\bar \Omega_\mathrm{GW}$ the
value in absence of stress, we have that $\Omega_\mathrm{GW} = D^2_f \bar \Omega_\mathrm{GW} $. In Fig. \ref{fig:D2f}
we show $D^2_f$ as a function of frequency.

\begin{center}
\begin{figure}[th]
\includegraphics[width=0.9\textwidth]{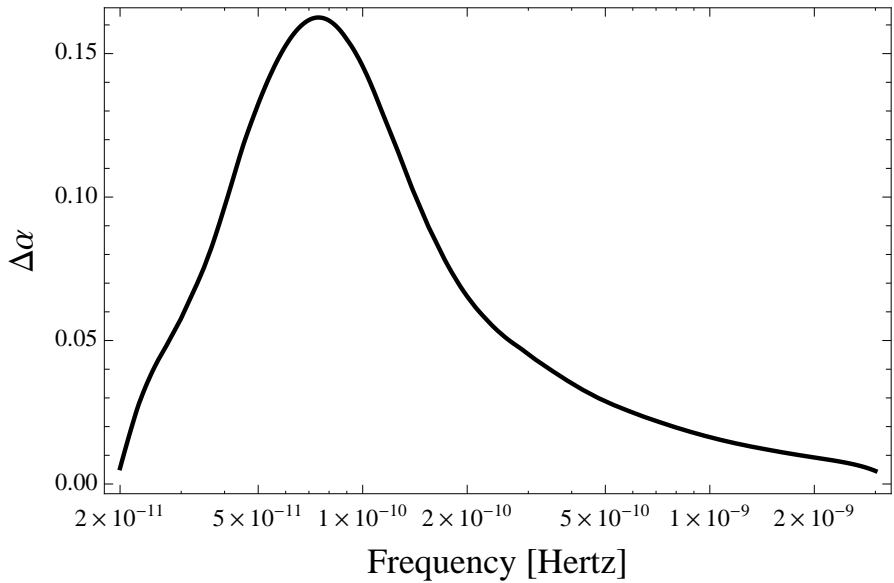}
\caption{Modification $\Delta\alpha$ to the logarithmic slope of a featureless power-law spectrum $f^\alpha$ at the source, introduced by the anisotropic stress.}
\label{fig:dalpha}
\end{figure}
\end{center}

The fact that the collisions introduce a frequency dependence on the amplitude of the GW background (in addition to any dependence that could be already present at the source) is a more interesting case with respect to a frequency independent suppression, since the latter could not be disentagled from
our ignorance on the amplitude of the original spectrum (that usually depends on the free parameters of the theory and thus cannot be determined \emph{a priori}
without any additional experimental input). On the other hand, a frequency-dependent effect can be disentangled by using this same dependence. 
To better quantify this, let us suppose that the GW spectrum at the source, has a featureless power law behaviour $\propto f^\alpha$. After correcting
for the expansion, the shape is still the same. Then the spectrum, once the effect of anisotropic stress has been taken into account, will be:
\begin{equation}
\Omega_{GW}(f)= A D^2_f f^\alpha.
\end{equation}
The logarithmic slope of the spectrum in a given point is given by $\alpha'=d\log\Omega_{GW}/d\log f$, so that, defining the deviation $\Delta\alpha=\alpha'-\alpha$, we have:
\begin{equation}
\Delta\alpha = \frac{d\log\Omega_\mathrm{GW}}{d\log f} - \alpha = \frac{d\log D^2_f}{d\log f}.
\end{equation}
In Fig. \ref{fig:dalpha}, we show $\Delta\alpha$ as a function of frequency. We find that it has a maximum value of $\Delta\alpha =0.15$ at $f\simeq 0.1\nHz$.

The nHz region of the spectrum of GWs can be probed by PTAs, combining the fact that pulsars are very stable clocks with the fact that the time of arrival of the pulse also depends on the GW background between the pulsar and the Earth \cite{PTA}. 
Correlating the electromagnetic signals from different pulsars could, in principle, enable a positive detection of the GW background in the galaxy. PTAs are sensitive to GWs with frequency between roughly 1 and  100 nHz, with a maximum sensitivity in the region between 3 and 10 nHz.
The lower bound comes from the fact that standard pulsar timing techniques absorb any low-frequency signal,
and so the time span of the data (currently $\sim 30$ years) gives a lower bound on the observable frequencies.
The principal source of GWs in the region of maximum sensitivity are believed to be coalescing supermassive binary black-hole
systems in the centre of merging galaxies. However, a signal of cosmological origin could also be present in that same frequency range,
for example GWs produced during the inflationary era or from the oscillation of cosmic string loops.

Can the effect of anisotropic stress leave an imprint that could be, in the future, detected using PTAs?
This could be possible, at some conditions. First of all, obviously, a signal of cosmological origin has to be present 
in the nHz range and it has to be strong enough to be detectable
by PTAs. As noted in the introduction, the oscillation of cosmic (super)strings loops could give a detectable signal in the nHz range. However,
GWs generated in this way are produced at scales smaller or equal than the horizon, so that the analysis made here does not rigorously apply. 
Although it is sensible to expect that GWs with wavelengths comparable to the horizon would behave in a similar way to the super-horizon GWs considered here,
we cannot assess the effect on the overall spectrum and we defer this generalization to a future work. This leaves the GWs produced during inflation
by the amplification of quantum vacuum fluctuations, that are too weak to be detected by ongoing PTA projects but could be within the reach of future experiments
like SKA. In any case, an independent observation of the signal at larger frequencies, like those probed by interferometers,
would also be useful to normalize the spectrum at  the source. Secondly, the cosmological signal
should be clearly separated from the astrophysical signal, like that produced by black hole binaries.
The third point is the more problematic. We have shown that the effects of the damping are more evident
at frequencies between 0.1 and 1 nHz. As explained above, the lower limit to the detectable frequency is
given by the time span of the observations, currenly 30 years. In 70 years from now, with a total of 100 years of observations, the lower
limit will be $f=1/(100 \mathrm{years})\simeq 0.3\nHz$, where the intensity of the wave is reduced by only 5\% and the deviation
from a featureless power law is $\Delta\alpha=0.15$. The maximum change of the slope is at $f\simeq 0.1\nHz$, corresponding
to 300 years of observations. It should also be noted that, even if we were willing to wait such a long time, it is not certain
that it would actually increase the sensitivity, because of intrinsic low-frequency instabilities in the timing data. In conclusion,
we think that the possibility to detect this effect are quite small, and that they will depend crucially on the existence of a sizeable inflationary-like GW background at the frequencies of interest and on the capability of estimating the logarithimic
slope of the spectrum in the region just below 1 nHz.

\ack
This work has been developed in the framework of the CGW collaboration
(www.cgwcollaboration.it).

\end{document}